\documentclass[aps,prd,twocolumn,showpacs,floats,floatfix,letterpaper,nofootinbib,superscriptaddress,]{revtex4}

\usepackage{amssymb,amsmath,latexsym,mathrsfs}
\usepackage{graphicx}
\usepackage{epsfig}
\usepackage{multirow}
\usepackage{array}
\usepackage{color}

\newcommand{\neff}{N_{\textrm{eff}}}
\newcommand{\cvis}{c^2_{\textrm{vis}}}
\newcommand{\ceff}{c^2_{\textrm{eff}}}

\newcommand{\mnu}{m_{\nu}}

\begin{document}

%%%%%%%%%%%%%%%%%%%%%%%%%%%%%%%%%%%%%%%%
%%%%%%%%%%%%%%%%%%%%%%%%%%%%%%%%%%%%%%%%

\title{Neutrino and Dark Radiation properties in light of latest CMB observations}
\author{Maria Archidiacono}
\affiliation{Department of Physics and Astronomy
University of Aarhus, DK-8000 Aarhus C, Denmark}
\author{Elena Giusarma}
\affiliation{IFIC, Universidad de Valencia-CSIC, 46071, Valencia, Spain}
\author{Alessandro Melchiorri}
\affiliation{Physics Department and INFN, Universita' di Roma 
	``La Sapienza'', Ple.\ Aldo Moro 2, 00185, Rome, Italy}
\author{Olga Mena}
\affiliation{IFIC, Universidad de Valencia-CSIC, 46071, Valencia, Spain}

\begin{abstract}

Recent Cosmic Microwave Background measurements at high multipoles from the
South Pole
Telescope and from the Atacama Cosmology Telescope  seem to disagree in
their conclusions for the neutrino and dark radiation properties. In this
paper we set new bounds on the dark radiation and neutrino properties in
different cosmological scenarios combining the ACT and SPT data with the
nine–year data release of the Wilkinson Microwave Anisotropy Probe
(WMAP-9), Baryon Acoustic Oscillation data, Hubble Telescope measurements
of the Hubble constant, and Supernovae Ia luminosity distance data.
In the standard three massive neutrino case, the two high multipole probes give similar results if Baryon Acoustic Oscillation data are removed from the analyses and
Hubble Telescope measurements are also exploited. A similar result is obtained within a standard cosmology with $\neff$ massless neutrinos,
although in this case the agreement between these two measurements is
also improved when considering simultaneously Baryon Acoustic Oscillation data
and Hubble Space Telescope measurements. In the $\neff$ massive
neutrino case the two high multipole probes give very different
results regardless of the external data sets used in the combined
analyses. 

When considering extended cosmological scenarios with a dark energy
equation of state or with a running of the scalar spectral index, the evidence for neutrino masses found for
the South Pole Telescope in the three neutrino scenario disappears for all the data
combinations explored here. Again, adding Hubble Telescope data seems
to improve the agreement between the two high multipole CMB
measurements considered here. In the case in which a dark radiation
background with unknown clustering properties is also considered, SPT 
data seem to exclude the standard value for the dark radiation viscosity
$\cvis=1/3$ at the $2\sigma$ CL, finding
evidence for massive neutrinos only when combining SPT data with BAO measurements.

\end{abstract}


\maketitle

%%%%%%%%%%%%%%%%%%%%%%%%%%%%%%%%%%%%%%%%
\section{Introduction}

Solar, atmospheric, reactor, and accelerator neutrinos have provided compelling evidence for the existence of neutrino oscillations. Barring exotic explanations, oscillation data imply non-zero neutrino masses. However, oscillation experiments only provide bounds on the neutrino mass squared differences, and therefore the measurement of the absolute scale of the neutrino mass must come from different observations.
In the Standard Model of elementary particles, there are three active
neutrinos. However, additional sterile neutrino species, or extra
relativistic degrees of freedom could also arise in a number of
extensions to the standard model of particle physics, as for instance, in axion
models~\cite{axions}, in  decaying of non-relativistic matter
models~\cite{decay}, in scenarios with gravity waves~\cite{gw}, extra
dimensions~\cite{extra}, early dark energy~\cite{ede} or in asymmetric
dark matter models~\cite{Blennow:2012de}. Cosmological data provide
a tool to test the neutrino properties, since the neutrino masses and
abundances affect both the Cosmic Microwave Background  (CMB) physics as
well as the galaxy clustering properties, see Refs~\cite{Mangano:2006ur,Hamann:2007pi,Acero:2008rh,Melchiorri:2008gq,Reid:2009nq,
Hamann:2010pw,rt,us,Hou:2011ec,Hamann,Hamann:2011hu,Nollett:2011aa,Giusarma,dePutter:2012sh,Joudaki,latest,Giusarma:2012ph,Archidiacono:2013xxa} for
constraints on the neutrino masses and/or abundances with
a variety of cosmological data sets and different assumptions
regarding the fiducial cosmology. 

On the other hand, cosmological measurements also
allow to test the clustering properties of the extra relativistic
degrees of freedom, parameterized via $\neff$, being $\neff=3.04$ in
the standard model scenario. The clustering pattern of the dark
radiation component is represented by its rest frame speed of sound
$\ceff$ and its viscosity parameter $\cvis$. The former parameter controls the
relationship between velocity and anisotropic stress, 
being these parameters  $\ceff=\cvis=1/3$ if  the dark radiation
background is composed by neutrinos. Several analyses have set bounds on these parameters~\cite{Diamanti:2012tg,Archidiacono:2012gv,Archidiacono:2011gq} under different assumptions regarding the underlying cosmological model.

Recently new CMB data have become available. The Wilkinson Microwave
Anisotropy Probe (WMAP) collaboration has presented the cosmological implications of their
final nine--year data release~\cite{Hinshaw:2012fq}, finding $\sum m_\nu <
0.44$~eV at the $95\%$~CL and $\neff=3.84 \pm 0.40$ (being $\neff$ the number of thermalised massless neutrino species) when they combine
their data with CMB small scale measurements (as those from previous
data releases from both the Atacama Cosmology Telescope ACT~\cite{Das:2011ak}
and the South Pole Telescope SPT ~\cite{spt09}), Baryon Acoustic
Oscillations (BAO) and Hubble Space Telescope (HST)
measurements. 

The SPT collaboration has also recently presented their
observations of 2540 deg$^2$ of sky, providing the CMB temperature
anisotropy power over the multipole range
$650<\ell<3000$~\cite{Hou:2012xq,Story:2012wx}, corresponding to the
region from the third to the ninth acoustic peak. The SPT
measurements have found evidence
for a decreasing power at high multipoles relative to the predictions
within a $\Lambda$CDM scenario, which suggest, potentially, that extensions
to the minimal $\Lambda$CDM scenario might be needed. In the case in
which massive neutrinos are added in the cosmological data analyses,
the SPT collaboration finds that the combination of SPT data with WMAP
(7 year data), together with Baryon Acoustic Oscillation (BAO) and
Hubble Space Telescope (HST) measurements shows a $2\sigma$ preference for these
models (when compared to the $\Lambda$CDM scenario). In the case of
three active massive neutrinos, they find $\sum m_\nu=0.32\pm 0.11$
after considering CMB, BAO, HST and SPT cluster measurements. However,
if the BAO measurements are removed and only CMB and HST data are
considered, the evidence for neutrino masses disappears at the
$95\%$~CL.  The authors of Ref.~\cite{Hou:2012xq} also find that when a
curvature component or a running in the spectral index of the
primordial perturbation spectrum are added as free parameters together
 with $\sum m_\nu$, the preference for nonzero neutrino masses is
 significantly reduced.  When $\neff$ massless neutrinos are
 considered, the bounds are $\neff=3.71\pm 0.35$ for the combination
 of CMB, BAO and HST data sets. Finally, when allowing for $\neff$
 massive neutrino species the bounds are $\sum m_\nu= 0.51 \pm
 0.15$~eV and $\neff=3.86\pm 0.37$, implying a $\sim 3\sigma$
  preference for $\sum m_\nu>0$ and a $2.2 \sigma$ preference for
  $\neff>3.046$. 

These findings, if confirmed by future CMB observations, as
those by the ongoing Planck mission~\cite{planck}, have an enormous impact
for Majorana neutrino searches. The mean value for $\sum m_\nu$ found by the SPT
collaboration implies a quasi-degenerate neutrino spectrum and
therefore the discovery of the neutrino character becomes at reach at
near future neutrinoless double beta decay experiments~\cite{GomezCadenas:2013ue}.

However, and also recently, the ACT collaboration has released new
measurements of the CMB damping tail~\cite{Sievers:2013wk}, finding a much
lower value for $\neff=2.79\pm 0.56$ when combining with WMAP 7 year
data. When considering also BAO and HST measurements, the value is
higher, $\neff=3.50\pm 0.42$.

The two data sets, SPT and ACT, seem also to disagree in the value of the lensing amplitude parameter
$A_L$ at more than $95\%$~CL~\cite{DiValentino:2013mt}. On the other
hand, ACT data do not seem to see evidence for neutrino masses,
 placing un upper limit of $\sum m_\nu<0.39$~eV at $95\%$~CL when ACT
data are combined with WMAP 7 year data together with BAO and HST measurements. 

We explore here the cosmological constraints in several
neutrino and dark radiation scenarios including the new WMAP 9 year data
as well as the new SPT and ACT measurements at high multipoles $\ell$. We also consider the
impact of other cosmological data sets, as BAO, HST and Supernova Ia
luminosity distance measurements.  We start with the massive
neutrino case within a $\Lambda$CDM scenario, setting bounds first on $\sum m_\nu$ assuming three
massive neutrinos and then moving to the case in which there are
$\neff$ massive species with a total mass given by $\sum m_\nu$.
We then enlarge the minimal $\Lambda$CDM scenario allowing for more
general models with a constant dark energy equation of state or with a
running of the scalar spectral index. We continue by studying the dark radiation properties, focusing first 
on the thermal abundances $\neff$ and adding after the dark radiation clustering
properties $\cvis$ and $\ceff$ as free parameters in the analysis.

The structure of the paper is as follows. In Sec.~\ref{sec:data} we
describe the data sets used in the numerical analyses as well as the
cosmological parameters used in each of the neutrino and dark
radiation models examined in Sec.~\ref{sec:analysis}. 
We draw our conclusions in Sec.~\ref{sec:concl}.

\section{Data and Cosmological parameters }
\label{sec:data}
The standard, three massive neutrino scenario we explore here is
described by the following set of parameters:
\begin{equation}
\label{parameter}
  \{\omega_b,\omega_c, \Theta_s, \tau, n_s, \log[10^{10}A_{s}], \sum m_\nu\}~,
\end{equation}
being $\omega_b\equiv\Omega_bh^{2}$ and $\omega_c\equiv\Omega_ch^{2}$  the physical baryon and cold dark matter energy densities,
$\Theta_{s}$ the ratio between the sound horizon and the angular
diameter distance at decoupling, $\tau$ is the reionization optical depth,
$n_s$ the scalar spectral index, $A_{s}$ the amplitude of the
primordial spectrum and $\sum m_\nu$ the sum of the masses of the
three active neutrinos in eV. We assume a degenerate neutrino mass
spectrum in the following.
The former scenario is enlarged with $\neff$ massive neutrinos in the case
of extended models
\begin{equation}
\label{parameter}
  \{\omega_b,\omega_c, \Theta_s, \tau, n_s, \log[10^{10}A_{s}], \neff,\sum m_\nu\}~,
\end{equation}
or with a constant dark energy equation of state $w$ (or with a
running of the scalar spectral index $n_{\textrm{run}}$) when
considering more general cosmological models:
\begin{equation}
\label{parameter}
  \{\omega_b,\omega_c, \Theta_s, \tau, n_s, \log[10^{10}A_{s}], w (n_{\textrm{run}}),
  \sum m_\nu\}~.
\end{equation}
We also study dark radiation models, described by $\Delta \neff$ relativistic
(i.e..massless) degrees of freedom together with three massive
neutrinos with $\sum m_\nu=0.3$~eV. This first dark
radiation scheme is described by
\begin{equation}
\label{parameter}
  \{\omega_b,\omega_c, \Theta_s, \tau, n_s, \log[10^{10}A_{s}], \Delta \neff\}~.
\end{equation}
Then we also consider extended parameter scenarios, with $\ceff$
and $\cvis$ also as free parameters:
\begin{equation}
\label{parameter}
  \{\omega_b,\omega_c, \Theta_s, \tau, n_s, \log[10^{10}A_{s}], \Delta \neff,
  \cvis, \ceff\}~,
\end{equation}
as well as the more general case in which the sum of the three neutrino
masses is also fitted to the data:
\begin{equation}
\label{parameter}
  \{\omega_b,\omega_c, \Theta_s, \tau, n_s, \log[10^{10}A_{s}], \Delta \neff,
  \cvis, \ceff, \sum m_\nu\}~,
\end{equation}
For our numerical analyses, we have used the Boltzmann CAMB code~\cite{camb} and extracted cosmological parameters from current data
using a Monte Carlo Markov Chain (MCMC) analysis based on the publicly available MCMC package \texttt{cosmomc}~\cite{Lewis:2002ah}.
Table \ref{tab:priors} specifies the priors considered on the different cosmological
parameters. Our neutrino mass prior is cast in the form of a (uniform)
prior on the neutrino density fraction $f_{\nu} =
\Omega_{\nu}/\Omega_{\rm DM}$, where $\Omega_{\nu}$ is the ratio of
the neutrino energy density over the critical density at redshift
zero, and $\Omega_{\rm DM}$ is the same ratio, but for the total dark matter density, which includes cold dark matter and neutrinos.

\begin{table}[h!]
\begin{center}
\begin{tabular}{c|c}
\hline\hline
 Parameter & Prior\\
\hline
$\Omega_{b}h^2$ & $0.005 \to 0.1$\\
$\Omega_{c}h^2$ & $0.01 \to 0.99$\\
$\Theta_s$ & $0.5 \to 10$\\
$\tau$ & $0.01 \to 0.8$\\
$n_{s}$ & $0.5 \to 1.5$\\
$\ln{(10^{10} A_{s})}$ & $2.7 \to 4$\\
$f_\nu$ &  $0 \to 0.2$\\
$\neff$ &  $1.047 \to 10$\\
&  ($0 \to 10$)\\
$w$ &  $-2 \to 0$\\
$n_{\textrm{run}}$ &  $-0.07 \to 0.02$\\
\hline\hline
\end{tabular}
\caption{Uniform priors for the cosmological parameters considered here.}
\label{tab:priors}
\end{center}
\end{table}

Our baseline data set is the nine--year WMAP data~\cite{Hinshaw:2012fq} (temperature and polarization)
with the routine for computing the likelihood supplied by the WMAP
team. We then also add CMB data from the SPT
experiment~\cite{Hou:2012xq,Story:2012wx}. In order to address for foreground contributions, the SZ amplitude $A_{SZ}$,
the amplitude of the clustered point source contribution, $A_C$, and
the amplitude of the Poisson distributed point source contribution,
$A_P$, are added as nuisance parameters in the CMB data
analyses. Separately, we also consider data from the ACT CMB
experiment~\cite{Sievers:2013wk}, in order to check the constraints on neutrino and
dark radiation properties with the combination of both WMAP plus SPT
data sets and WMAP plus ACT data sets. 
To the CMB basic data sets we add the latest constraint on the Hubble constant $H_0$ from the Hubble Space Telescope (HST)~\cite{Riess:2011yx}, or supernova data from
the 3 year Supernova Legacy Survey (SNLS3), see Ref.~\cite{snls3}. We
do not consider the combination of HST and SNLS3 measurements  because
these two data sets are not totally independent. In the case of SNLS3 data, we add in the MCMC analysis two
extra nuisance parameters related to the light curve fitting procedure used to analyse the supernova (SN) data.
These parameters characterise the dependence of the intrinsic
supernova magnitude on stretch (which measures the shape of the SN
light curve) and color~\cite{snls3}. 
Galaxy clustering measurements are considered in our analyses via
BAO signals. We use here the BAO signal from DR9~\cite{anderson} of the Baryon Acoustic Spectroscopic Survey (BOSS)~\cite{boss, boss2012}, with a median
redshift of $z=0.57$. Together with the CMASS DR9 data, we also include the recent measurement of the BAO scale based on a re-analysis (using reconstruction \cite{eisetal07})
of the LRG sample from Data Release 7 with a median redshift of
$z=0.35$~\cite{nikhil}, the measurement of the BAO signal at a lower
redshift $z=0.106$ from the 6dF Galaxy Survey 6dFGS~\cite{6dFGS} and
the BAO measurements from the WiggleZ Survey at $z=0.44$, $z=0.6$ and
$z=0.73$~\cite{Blake:2011en}. The data combinations for which we will
show results in the next section are the following: WMAP and SPT/ACT;
WMAP, SPT/ACT and HST; WMAP, SPT/ACT and SNLS3; WMAP, SPT/ACT and BAO;
WMAP, SPT/ACT, HST and BAO; and finally WMAP, SPT/ACT, SNLS3 and BAO.

%%%%%%%%%%%%%%%%%%%%%%%%%%%%%%%%%%%%%%%%
\section{Results}
\label{sec:analysis}
Here we present the constraints from current cosmological data sets on
the neutrino thermal abundance $\neff$ and on the sum of their masses
$\sum m_\nu$ in different scenarios, considering separately SPT and
ACT CMB data sets.

\subsection{Standard Cosmology plus massive neutrinos}
\label{sec:st}
\begin{figure*}
\begin{tabular}{c c}
\includegraphics[width=9cm]{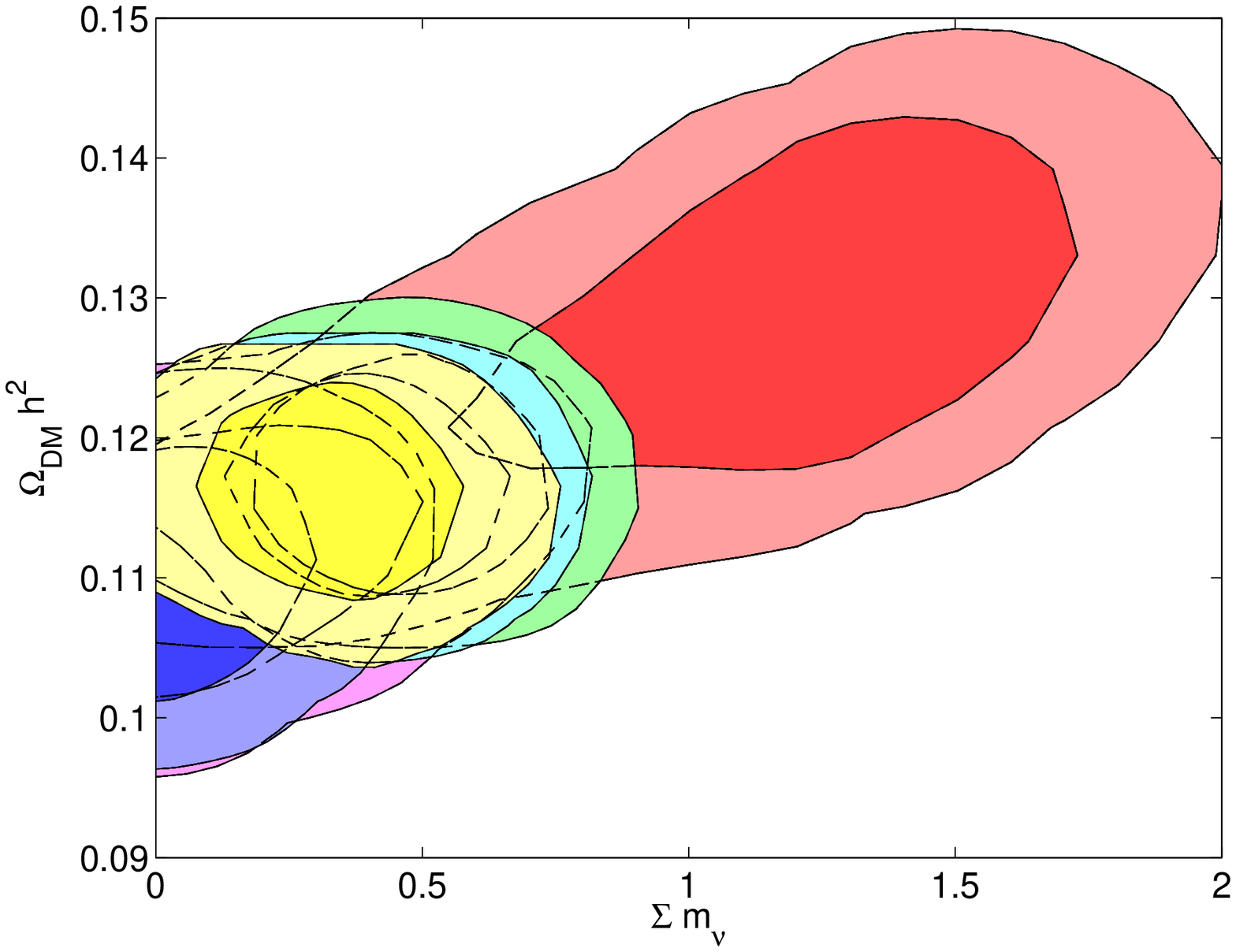}&\includegraphics[width=9cm]{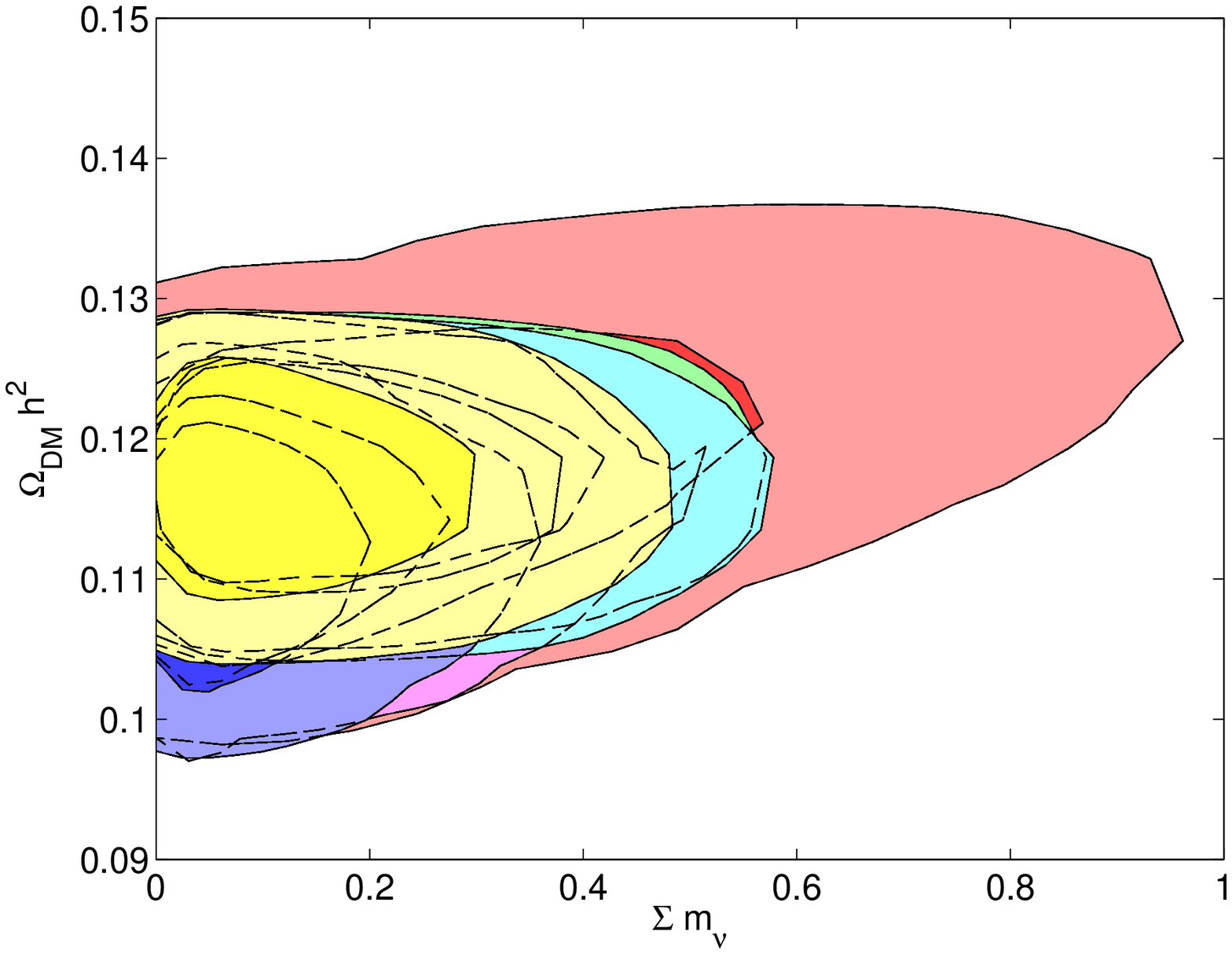}\\
\end{tabular}
 \caption{\small{Left panel (Three massive neutrino case): the red contours show the $68\%$ and $95\%$~CL allowed
  regions  from the combination of
  WMAP and SPT measurements in the ($\sum
  m_\nu$ (eV), $\Omega_{\textrm{dm}}h^2$) plane, while the magenta (blue) ones show the impact of
  the addition of SNLS3 (HST) data sets.  The green contours depict
  the results from the combination of CMB and BAO data, while the cyan and
  yellow ones show the impact of the SNLS3 (HST) data combined
  with CMB and BAO measurements. Right panel: as in the left panel but considering ACT data
  instead of SPT.}}
\label{fig:standardspt}
\end{figure*}

\begin{figure*}
\begin{tabular}{c c}
\includegraphics[width=9cm]{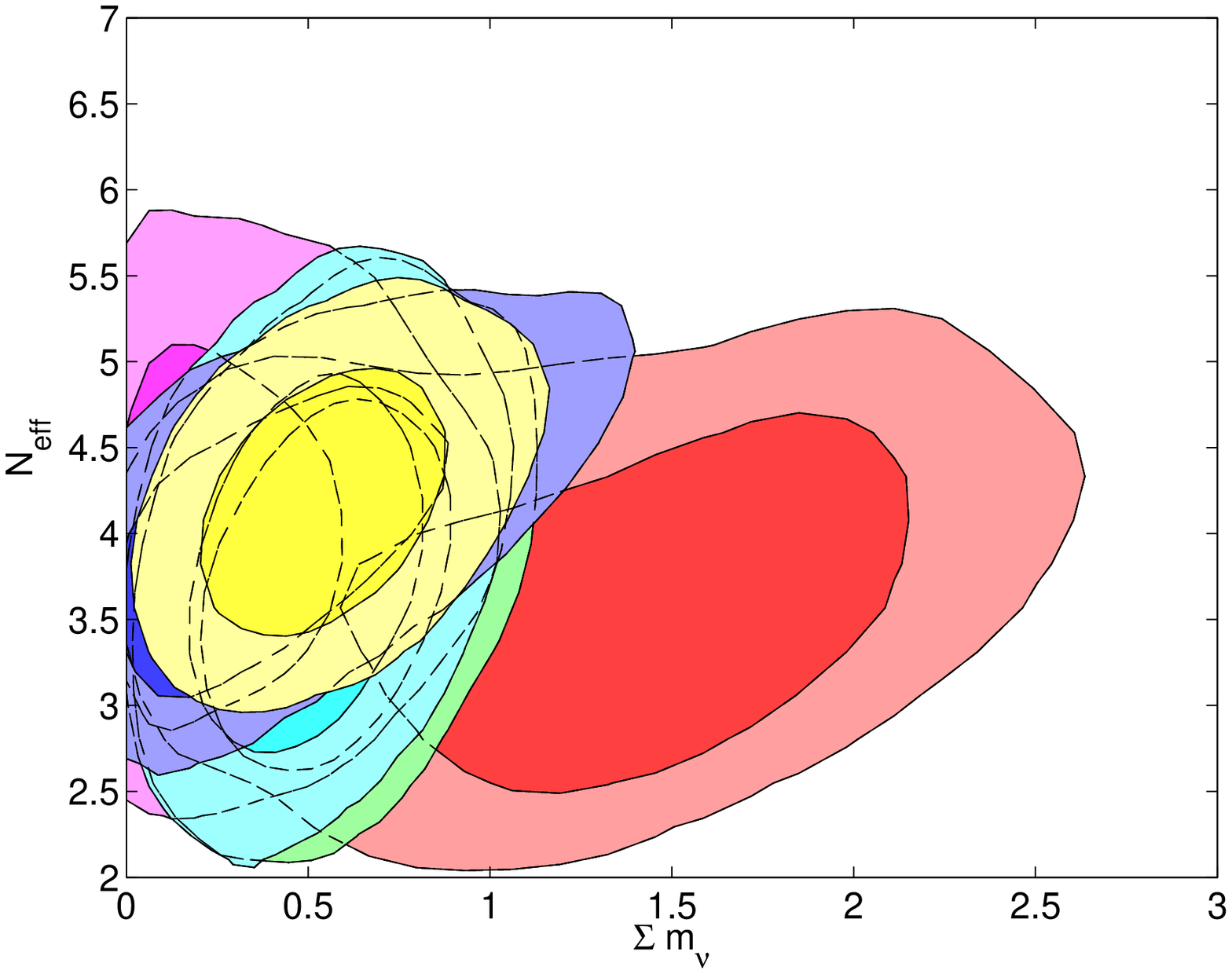}&\includegraphics[width=9cm]{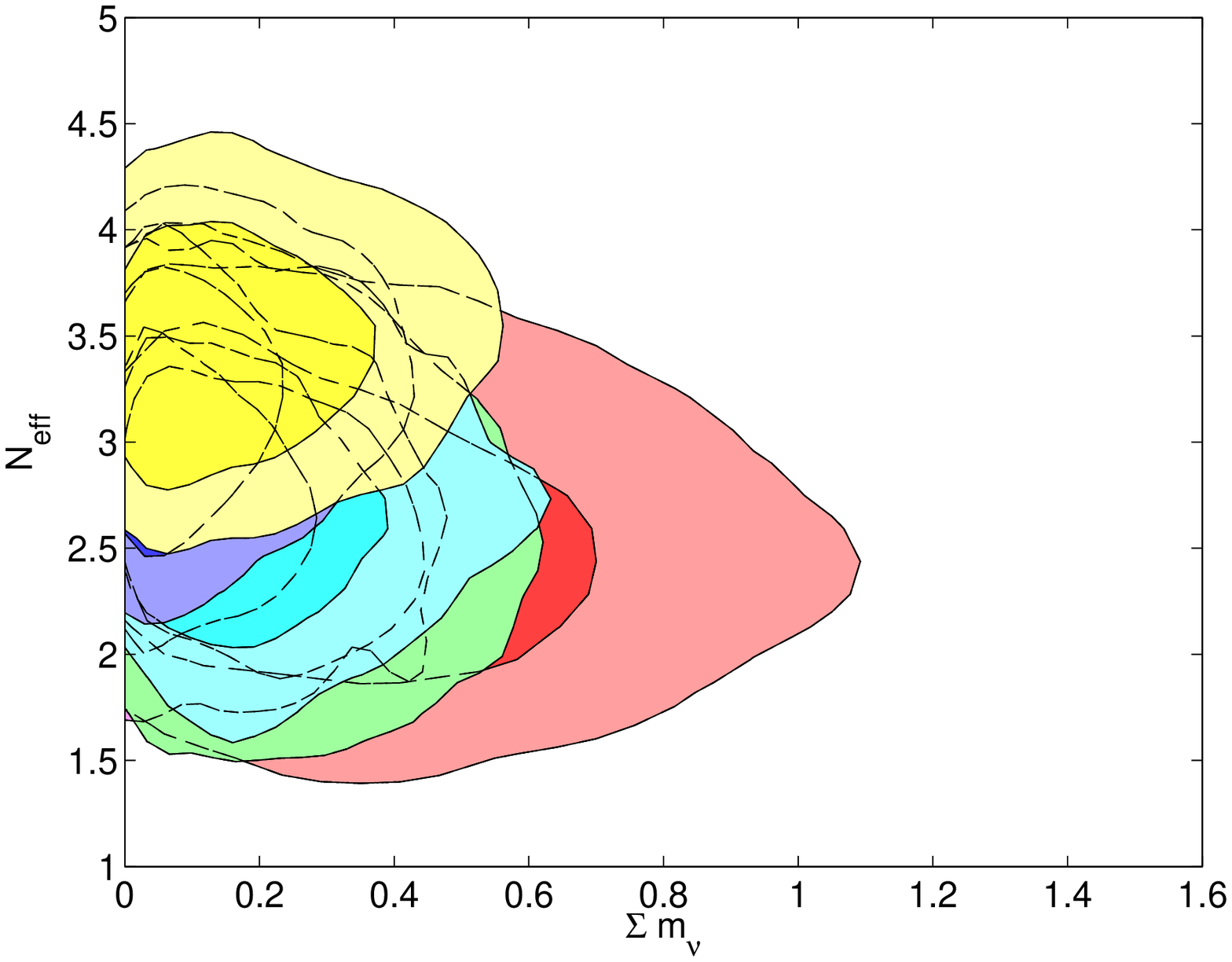}\\
\end{tabular}
 \caption{\small{Left panel ($\neff$ massive neutrino case): the red contours show the $68\%$ and $95\%$~CL allowed
  regions  from the combination of
  WMAP and ACT measurements in the ($\sum
  m_\nu$ (eV), $\neff$) plane, while the magenta (blue) ones show the impact of
  the addition of SNLS3 (HST) data sets.  The green contours depict
  the results from the combination of CMB and BAO data, while the cyan and
  yellow ones show the impact of the SNLS3 (HST) data combined
  with CMB and BAO measurements. Right panel: as in the left panel but
  considering ACT data instead of SPT.}}
\label{fig:standardact}
\end{figure*}

Through this section we shall assume a $\Lambda$CDM cosmology with
either three or $\neff$ light massive neutrinos. 
The left panels of Figs.~\ref{fig:standardspt} and \ref{fig:standardact}, depict our results for the
three and $\neff$ massive neutrino assumptions, respectively, in the case of considering
SPT CMB data, combined with the other data sets exploited here.
 Tables~\ref{tab:c1} and \ref{tab:c2} present the mean values and
 errors (or $95\%$ CL bounds) in the three and $\neff$ massive
 neutrino scenarios in the case of considering SPT  for the different data combinations detailed in the previous section. Our
results agree with those presented in Ref.~\cite{Hou:2012xq} by the
SPT collaboration. Notice that BAO data are crucial for the preference for
massive neutrinos in the three massive neutrino case, in which
 $\sum m_\nu=0.33 \pm 0.17 $ ($\sum m_\nu=0.40\pm 0.18$~eV) for
CMB plus BAO plus HST (SNLS3) data. In the $\neff$
massive  neutrino scenario, the bounds are $\sum m_\nu=0.56 \pm 0.23$~eV and
$\neff=4.21\pm 0.46$ ($\sum m_\nu=0.50\pm 0.21$~eV and $\neff=3.87\pm
0.68$)  for CMB plus BAO plus HST (SNLS3) data. 

If BAO data are removed, the preference for massive neutrinos disappears
in the three massive neutrino case, with a $95\%$~CL upper limit on the sum
of the three active neutrinos of  $\sum m_\nu <0.50$~eV in the case of
considering WMAP, SPT and HST measurements. For the same combination
of data sets, in the $\neff$ massive neutrino case explored here,
$\sum m_\nu = 0.48\pm 0.33$~eV and $\neff=4.08\pm 0.54$.  

We then consider separately new ACT data and perform identical
analyses to the ones done with SPT data, see Tabs.~\ref{tab:c3}, \ref{tab:c4}. Figures
~\ref{fig:standardspt} and \ref{fig:standardact} (right panels) depict our results for the
three and $\neff$ massive neutrino assumptions, respectively, in the case of considering
ACT CMB data combined with the other data sets described in the
previous section. Notice that there is no evidence for neutrino masses
in any of the data combinations explored here. 
A $95\%$~CL upper limit on the sum of the neutrino
masses of $\sum m_\nu <0.44$~eV ($<0.54$~eV) is found when considering CMB,  BAO and HST
(SNLS3) data, which agrees with the results presented in
Ref.~\cite{Sievers:2013wk}. In the $\neff$ massive neutrino case, we find
$\sum m_\nu<0.50$~eV ($\sum m_\nu<0.53$~eV) at $95\%$~CL and
$\neff=3.44\pm 0.37$ ($\neff=2.77\pm 0.46$) when considering CMB,  BAO and HST
(SNLS3) data. Only when adding HST measurements the allowed values of
$\neff$ are larger than $3$, see Tab.~\ref{tab:c4}, bringing the mean
value of $\neff$ closer to the one found in the SPT data analyses. 
When removing BAO data, we get  $\sum m_\nu<0.34$~eV ($95\%$~CL) for the
combination of CMB and HST measurements in the three massive neutrino
case and $\sum m_\nu<0.39$~eV  ($95\%$~CL), $\neff<3.20\pm 0.38$ in
the $\neff$ massive neutrino case. 

Therefore, we conclude that, within a standard cosmology with three massive neutrinos, ACT and SPT CMB measurements are compatible
if BAO data are not considered in the analyses and if a prior on $H_0$
from the HST experiment is also considered. However, the predictions in the $\neff$ massive neutrino case arising from ACT and SPT data
are not consistent even if BAO data are removed and a prior on $H_0$ from the
HST experiment is also added. 

\begin{table*}
\begin{center}
\begin{tabular}{lccccccc}
\hline \hline
            &      W9+SPT & W9+SPT & W9+SPT & W9+SPT& W9+SPT & W9+SPT  \\
      &                      & + HST    &+BAO     &+SNLS3 & +BAO+HST&+BAO+SNLS3\\      
\hline
\hspace{1mm}\\
               
$\sum \mnu$ (eV)& $1.14\pm0.41$  & $<0.50$  & $0.46\pm0.18$  & $<0.80$  & $0.33\pm0.17$ & $0.40\pm0.18$  \\
                \hline
\hline
\end{tabular}
\caption{Mean values and errors (or $95\%$~CL upper bounds) on $\sum
  \mnu$ (in eV) in a standard cosmology with three massive neutrinos for the
  different combinations of data sets in the case of considering SPT high multipole data.}
\label{tab:c1}
\end{center}
\end{table*}

\begin{table*}
\begin{center}
\begin{tabular}{lccccccc}
\hline \hline
   &             W9+SPT & W9+SPT & W9+SPT & W9+SPT& W9+SPT & W9+SPT  \\
   &                            & + HST    &+BAO     &+SNLS3 & +BAO+HST&+BAO+SNLS3\\
\hline
\hspace{1mm}\\
${\neff}$   & $3.66\pm0.61$ & $4.08\pm0.54$ & $3.76\pm0.67$ & $4.04\pm
0.68$ & $4.21\pm0.46$ & $3.87\pm0.68$\\
\hspace{1mm}\\ 
$\sum \mnu$ (eV)  & $1.35\pm0.55$ & $0.48\pm0.33$ & $0.56\pm0.22$ & $<0.91$ & $0.56\pm0.23$ & $0.50\pm0.21$  \\
\hline
\hline
\end{tabular}
\caption{Mean values and errors(or $95\%$~CL bounds)  on ${\neff}$ and
  $\sum \mnu$ (in eV) in a
  standard cosmology with $\neff$ massive neutrinos for the
  different combinations of data sets in the case of considering SPT high
  multipole data.}
\label{tab:c2}
\end{center}
\end{table*}

\begin{table*}
\begin{center}
\begin{tabular}{lccccccc}
\hline \hline
            &      W9+ACT & W9+ACT & W9+ACT & W9+ACT& W9+ACT & W9+ACT  \\
      &                      & + HST    &+BAO     &+SNLS3 & +BAO+HST&+BAO+SNLS3\\      
\hline
\hspace{1mm}\\
               
$\sum \mnu$ (eV) & $<0.89$  & $<0.34$  & $<0.53$  & $<0.49$  & $<0.44$ & $<0.54$  \\
                
\hline
\hline
\end{tabular}
\caption{$95\%$~CL upper bounds on $\sum \mnu$ (in eV) in a standard cosmology with three massive neutrinos for the
  different combinations of data sets in the case of considering ACT high
  multipole data.}
\label{tab:c3}
\end{center}
\end{table*}

\begin{table*}
\begin{center}
\begin{tabular}{lccccccc}
\hline \hline
   &             W9+ACT & W9+ACT & W9+ACT & W9+ACT& W9+ACT & W9+ACT  \\
   &                            & + HST    &+BAO     &+SNLS3 & +BAO+HST&+BAO+SNLS3\\
\hline
\hspace{1mm}\\
${\neff}$   & $2.64\pm0.51$ & $3.20\pm0.38$ & $2.63\pm0.48$ & $2.75\pm0.44$ & $3.44\pm037$ & $2.78\pm0.46$ \\
\hspace{1mm}\\
$\sum \mnu$ (eV)  & $<0.95$ & $<0.39$ & $<0.55$ & $<0.44$ & $<0.50$ & $<0.53$  \\
\hline
\hline
\end{tabular}
\caption{Mean values and errors on ${\neff}$ and $95\%$~CL upper
  bounds on $\sum \mnu$ (in eV) in a standard cosmology with $\neff$ massive neutrinos for the
  different combinations of data sets in the case of considering ACT high
  multipole data.}
\label{tab:c4}
\end{center}
\end{table*}

\subsection{Massive neutrinos and extended cosmologies}

In this section we compute the bounds on the sum of the three active
neutrino masses considering extended cosmologies with a dark energy
equation of state or with a running of the scalar spectral index.

\begin{figure*}
\begin{tabular}{c c}
\includegraphics[width=9cm]{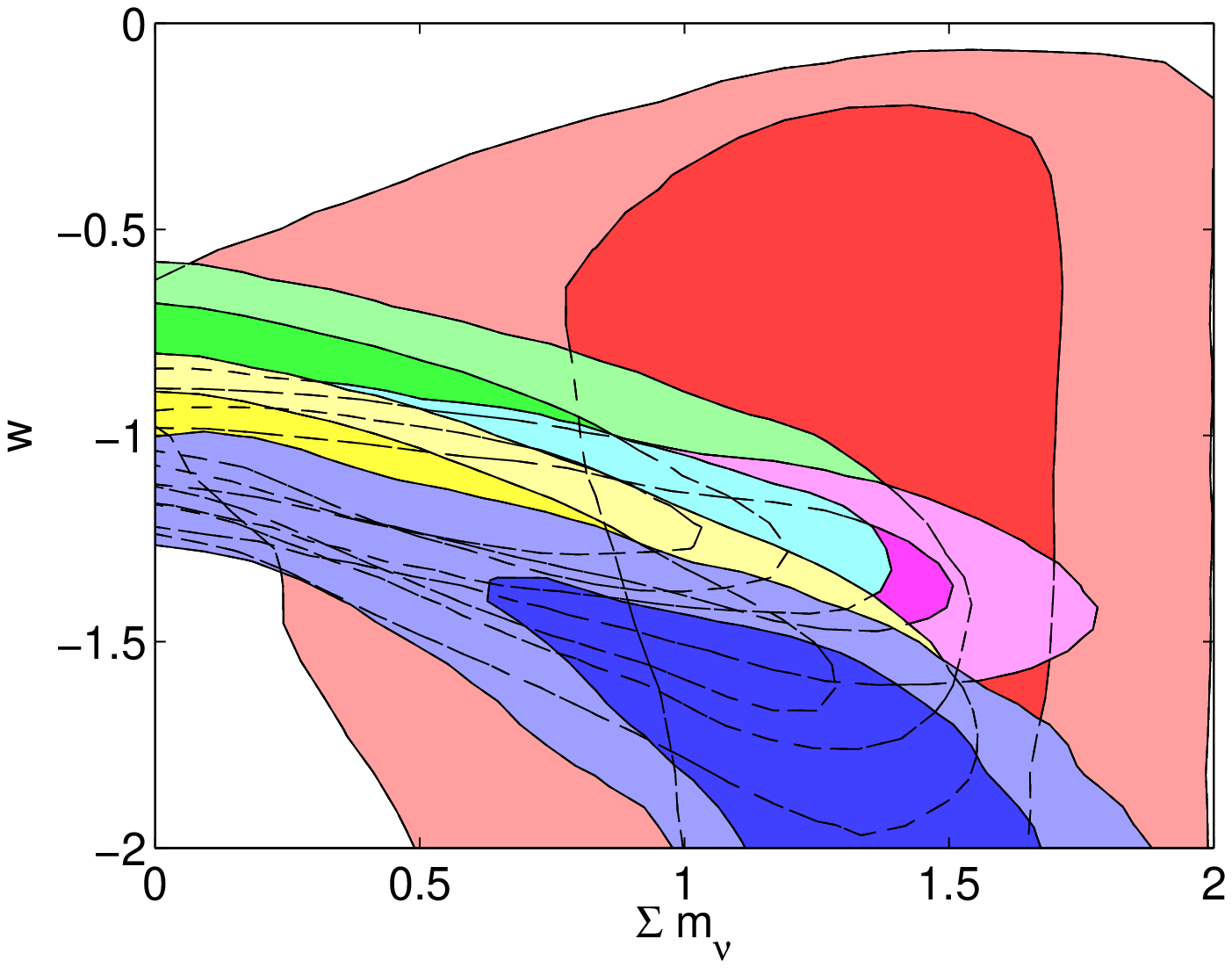}&\includegraphics[width=9cm]{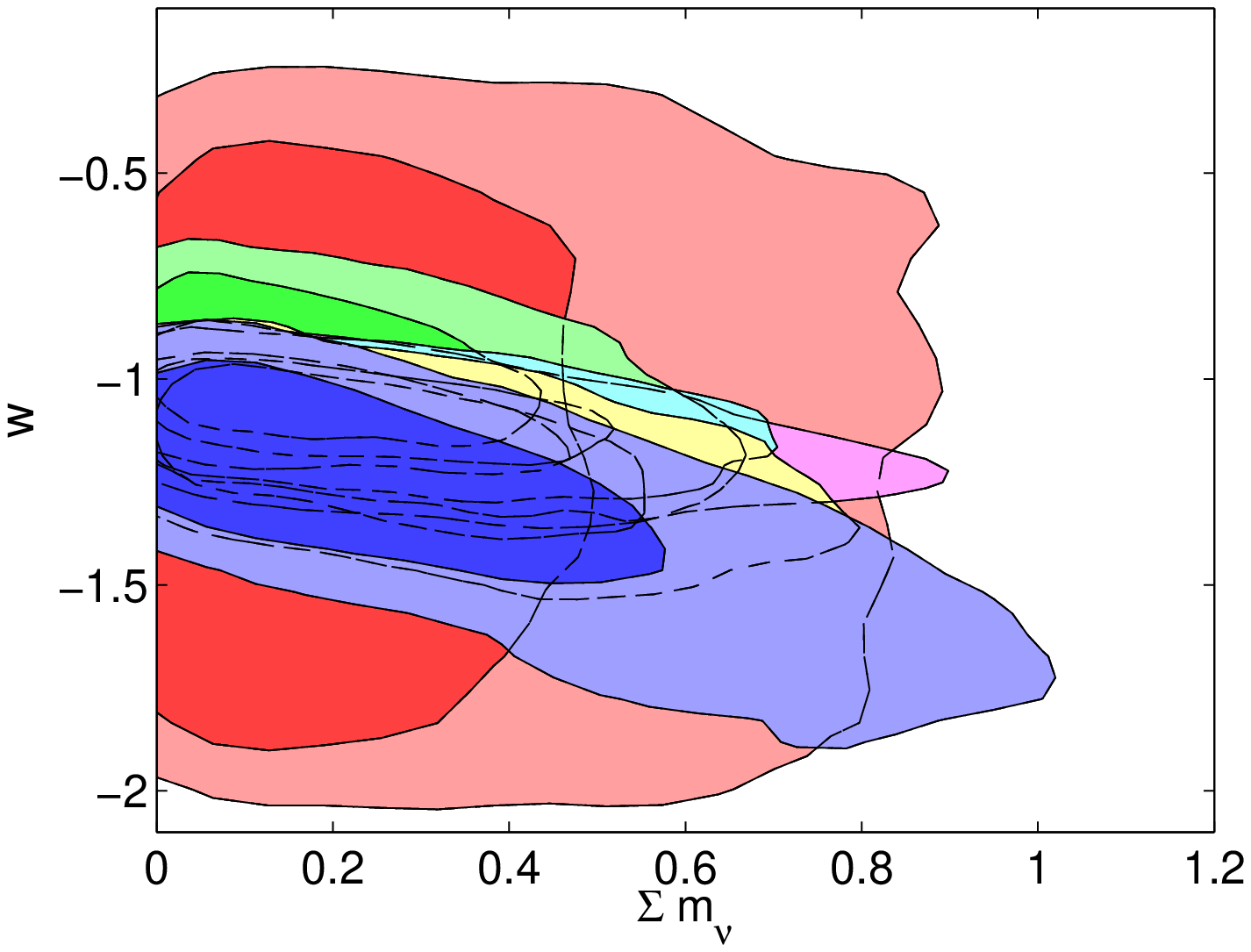}\\
\end{tabular}
 \caption{\small{Left panel (Three massive neutrino case plus dark energy): the red contours show the $68\%$ and $95\%$~CL allowed
  regions  from the combination of
  WMAP and SPT measurements in the ($\sum
  m_\nu$ (eV), $w$) plane, while the magenta (blue) ones show the impact of the addition of SNLS3 (HST) data sets.  The green contours depict
  the results from the combination of CMB and BAO data, while the cyan and
  yellow ones show the impact of the SNLS3 (HST) data combined
  with CMB and BAO measurements. Right panel: as in the left panel but
  for the case of ACT data.}}
\label{fig:w}
\end{figure*}

Concerning the dark energy equation of state $w$, there is a strong and very well known degeneracy among the sum of
neutrino masses and the dark energy equation of state $w$, see Ref.~\cite{Hannestad:2005gj}. The bounds from cosmology on the sum of the neutrino
masses will be much weaker if the dark energy fluid is not interpreted
as a cosmological contant, in which case, the dark energy equation of
state will be an extra free parameter. If $w$ is allowed to vary,
$\Omega_{\rm dm}$ can be much higher and consequently the neutrino
mass also increases to leave unchanged the matter power spectrum and
the growth of matter perturbations. The SPT collaboration~\cite{Hou:2012xq} has also
considered the impact of a constant dark energy equation of state $w$, and
they find $\sum m_\nu =0.27\pm 0.11$~eV for the combination of their CMB and clusters data with
WMAP 7 year, HST and BAO data sets. Figure~\ref{fig:w}, left panel,
shows our results for SPT data within the different combinations of
data sets addressed here. Notice that, in general, the evidence for
neutrino masses in much milder than in the cosmological constant
case, and the bounds on $\sum m_\nu$ are much larger than those shown in
Tab.~\ref{tab:c1} due to the degeneracy between $\sum m_\nu$ and $w$. 
 Supernovae measurements are, for this particular case, more
useful than the $H_0$ prior from the HST experiment. 
Figure~\ref{fig:w}, right panel, shows the constraints in the ($\sum
  m_\nu$ (eV), $w$)  plane in the case of considering ACT data. Notice
  that the bounds on $\sum m_\nu$ are tighter than those found for the
  case of analysing SPT data. Indeed, the bounds on the sum of the
  three massive neutrino masses computed for the case of a dark energy
  equation of state $w\neq -1$ are not very different from those
  obtained for a $\Lambda$CDM universe, see Tab.~\ref{tab:c3}. 
SNLS3 measurements  have a much larger constraining power than the HST
prior also in the ACT data analyses performed in this section, especially
  for measuring the dark energy equation of state $w$. 

\begin{figure*}
\begin{tabular}{c c}
\includegraphics[width=9cm]{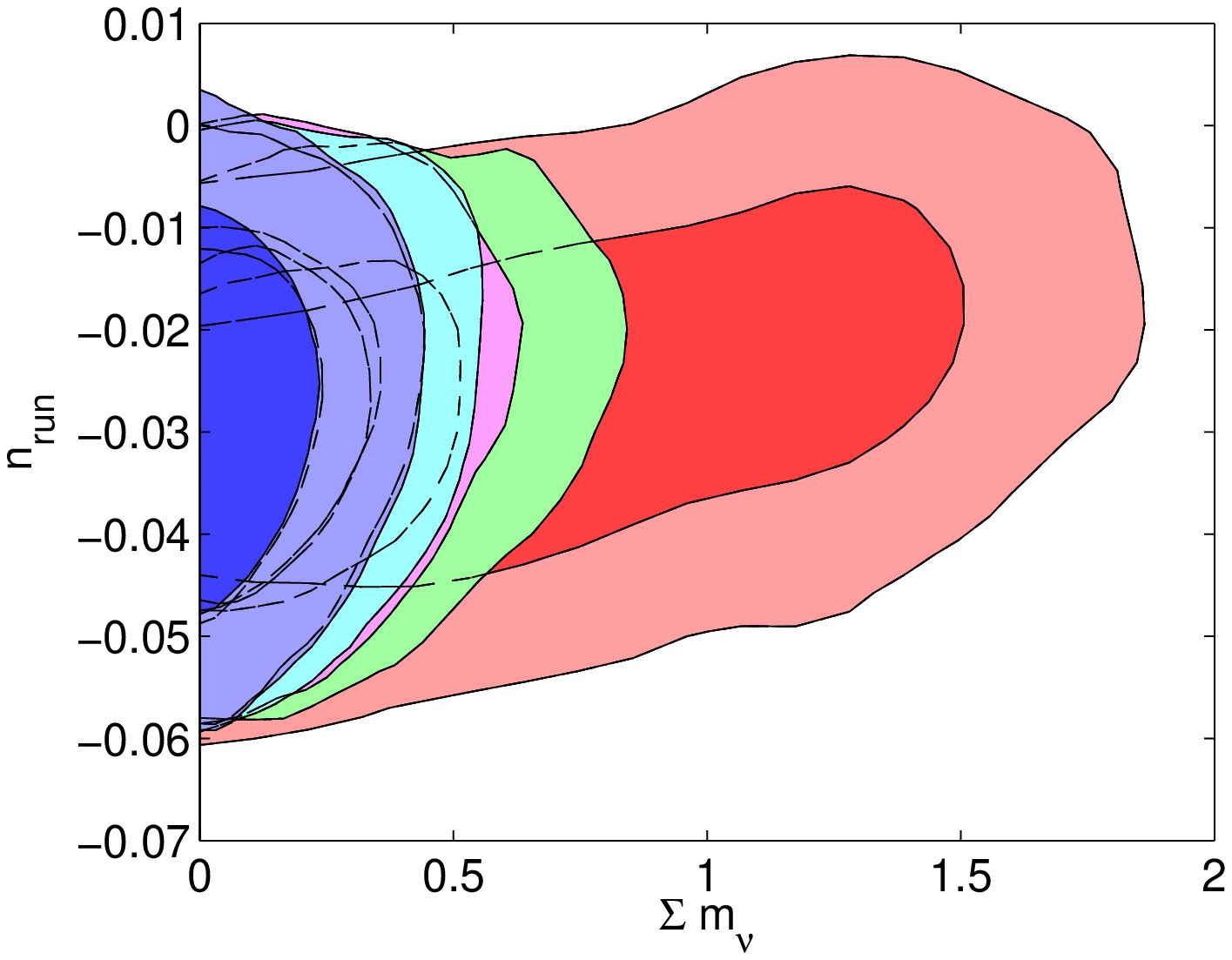}&\includegraphics[width=9cm]{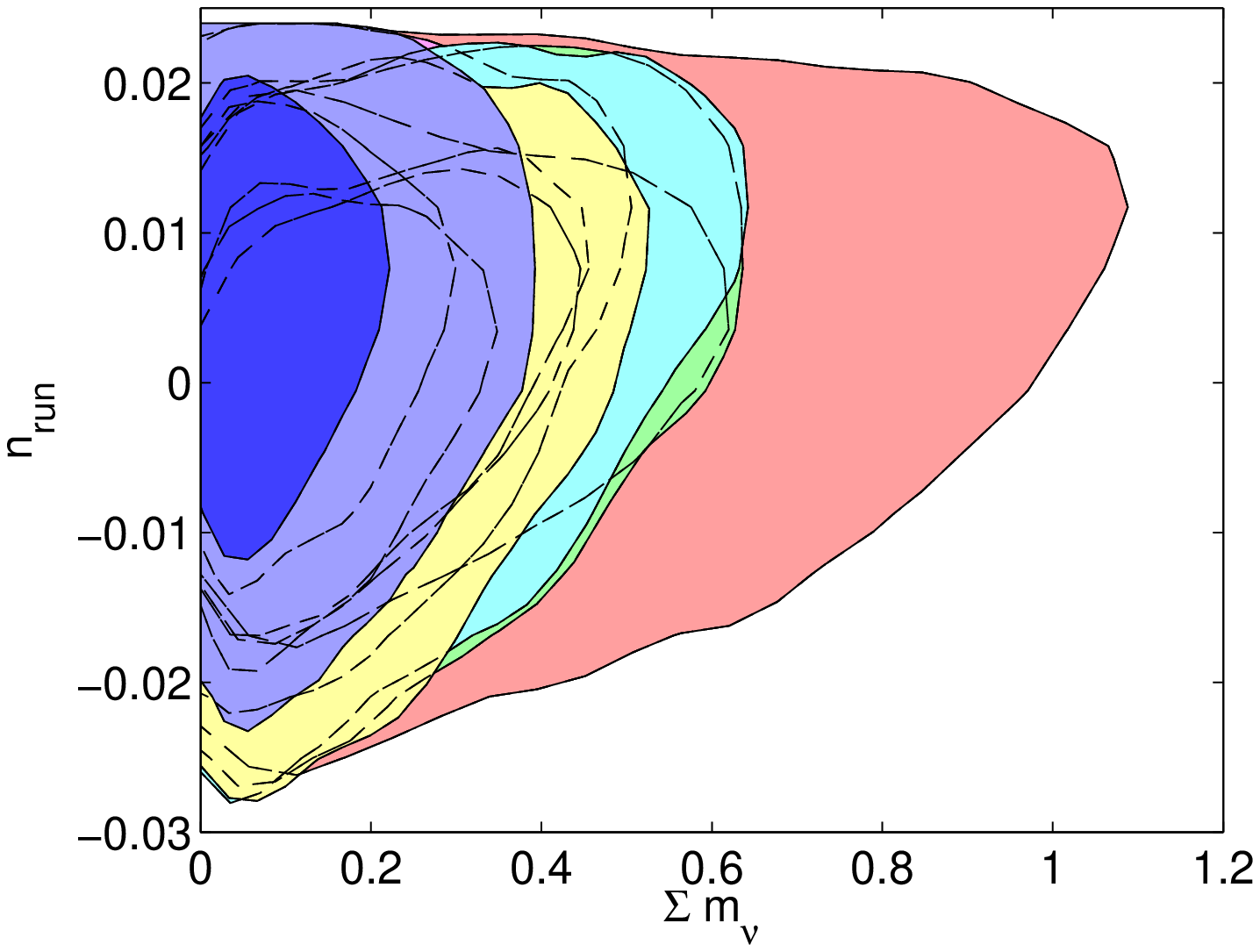}\\
\end{tabular}
 \caption{\small{Left panel (Three massive neutrino case plus $n_{\textrm{run}})$: the red contours show the $68\%$ and $95\%$~CL allowed
  regions  from the combination of
  WMAP and SPT measurements in the ($\sum
  m_\nu$ (eV), $n_{\textrm{run}}$) plane, while the magenta (blue)
  ones show the impact of the addition of SNLS3 (HST) data sets. 
 The green contours depict the results from the combination of CMB and
 BAO data, while the cyan and
  yellow ones show the impact of the SNLS3 (HST) data combined
  with CMB and BAO measurements. Right panel: as in the left panel but
  for the case of ACT data.}}
\label{fig:nrun}
\end{figure*}

We also explore the case in which a running in the spectral index of
primordial perturbations is added to the minimal $\Lambda$CDM
cosmology.  In general, the spectrum of the scalar perturbations is
not exactly a power law but it varies with scale.
 Therefore one must consider the scale dependent running of the spectral index $n_{\rm run}\:=\:dn_s/d\ln k$.
Following \cite{Kosowsky:1995aa}, the power spectrum for the scalar perturbations reads
\begin{equation*}
P(k) \equiv A_sk^{n(k)} \propto \left( k\over k_0 \right)^{n_s\, +\, \ln(k/k_0)
(dn/d\ln k)\, +\, \cdots }~,
\end{equation*}
being $k_0=0.002$~Mpc$^{-1}$ the pivot scale. 
Figure~\ref{fig:nrun}, left panel, shows our results for SPT data within the different combinations of
data sets addressed here. The evidence for
neutrino masses found for the SPT data in the cosmological constant
case disappears in all the data combinations explored
here. We find,  for the case of the SPT data analyses, a $2\sigma$
preference for a negative running, in agreement with the results presented in
Ref.~\cite{Hou:2012xq}.

Figure~\ref{fig:nrun}, right panel, shows our results for ACT data in
the case of considering a running in the scalar spectral index. The
bounds on the sum of the three massive neutrinos are now very similar
to those found for the SPT experiment and also very similar to those
found for ACT in the case of the minimal $\Lambda$CDM scenario. 
However, the preferred region for $n_{\textrm{run}}$ is perfectly
consistent with no running of the scalar spectral index, in agreement
with the results presented by the ACT team~\cite{Sievers:2013wk}.

\subsection{Standard cosmology plus dark radiation}
\label{sec:stdark}

\begin{figure*}
\begin{tabular}{c c}
\includegraphics[width=9cm]{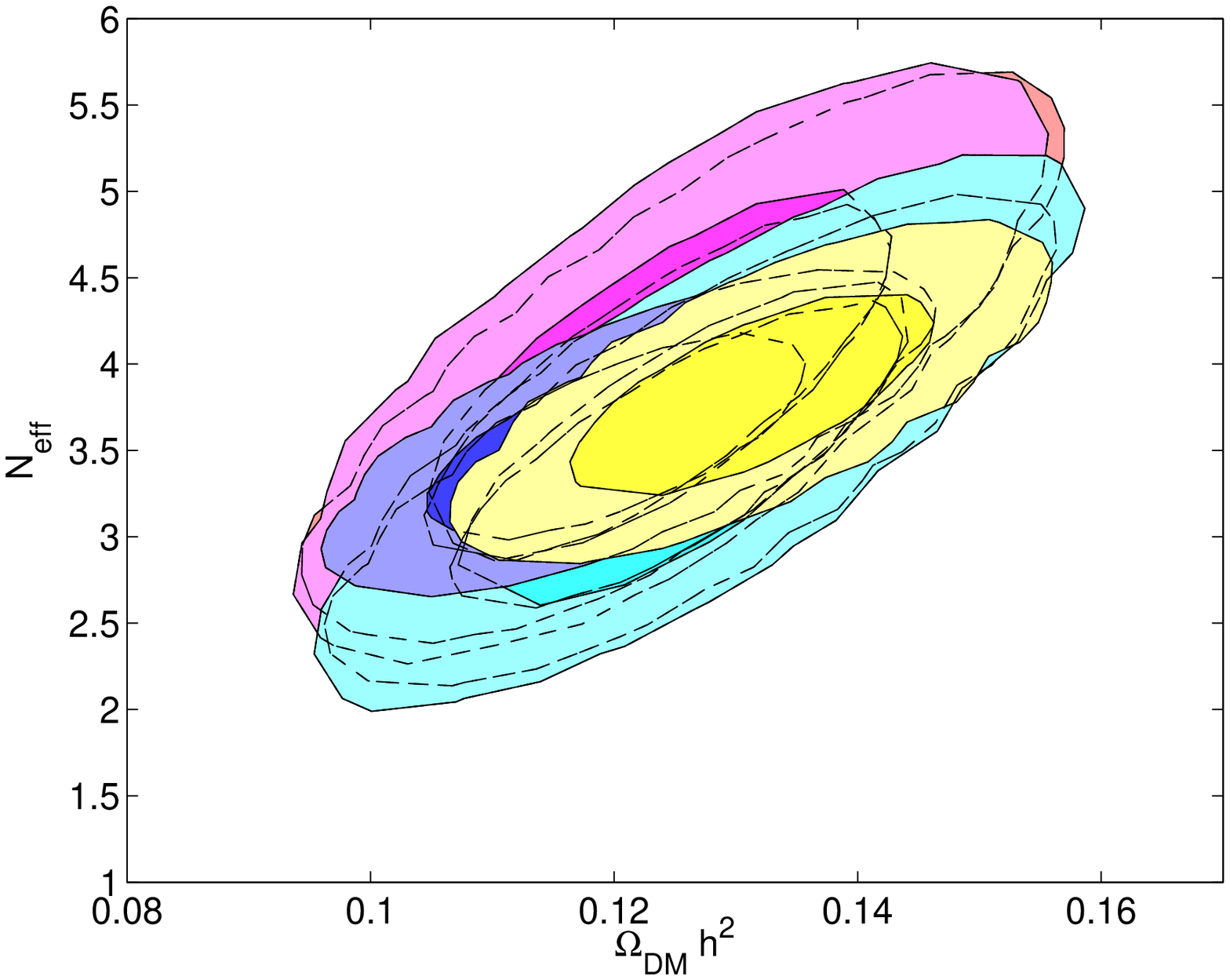}&\includegraphics[width=9cm]{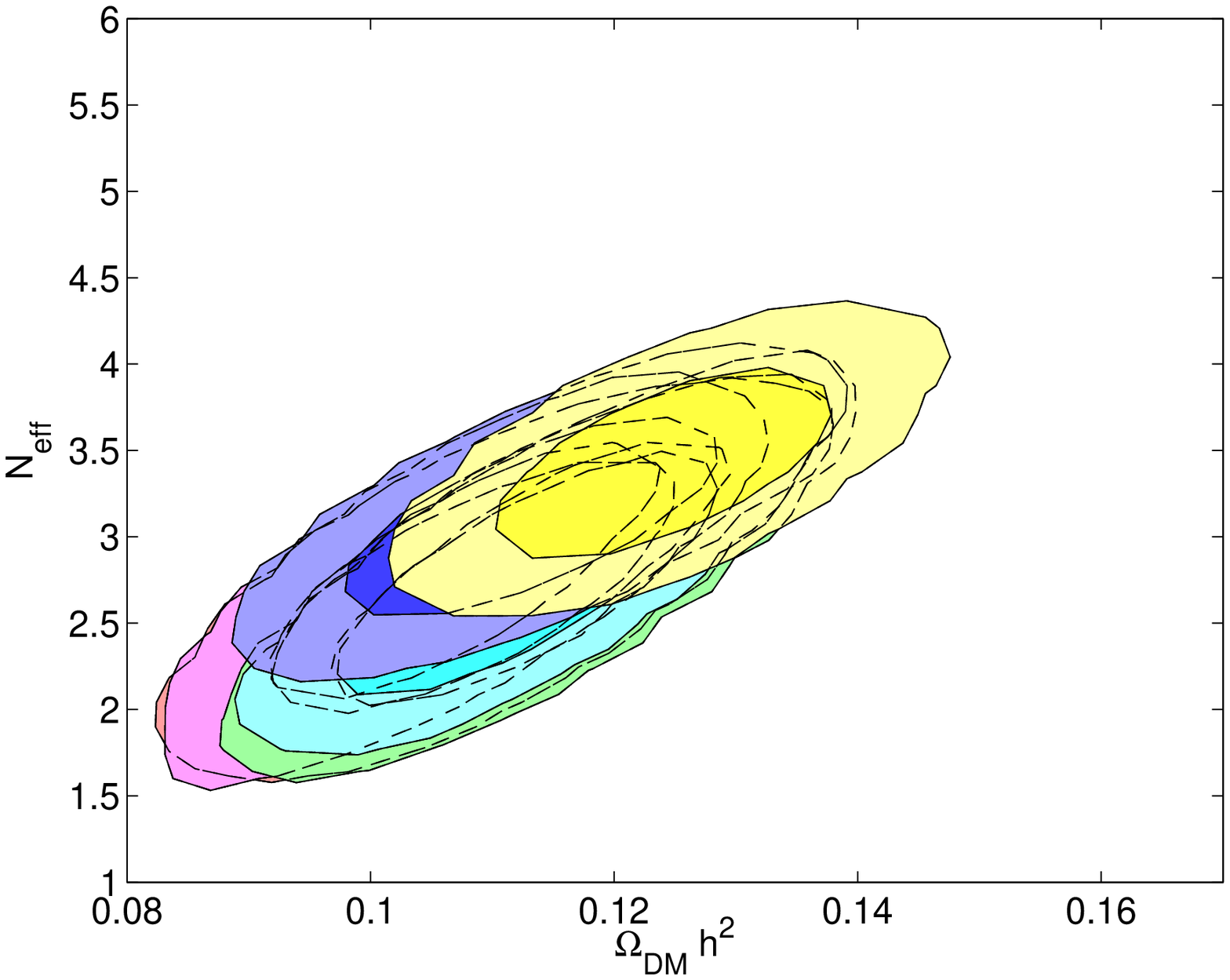}\\
\end{tabular}
 \caption{\small{Left panel (Massless neutrino case): the red contours show the $68\%$ and $95\%$~CL allowed
  regions  from the combination of
  WMAP and SPT measurements in the ($\Omega_{\textrm{dm}}h^2$, $\neff$) plane, while the magenta (blue) ones show the impact of
  the addition of SNLS3 (HST) data sets.  The green contours depict
  the results from the combination of CMB and BAO data, while the cyan and
  yellow ones show the impact of the SNLS3 (HST) data combined
  with CMB and BAO measurements. Right panel: as in the left panel but
  considering ACT data instead of SPT CMB data.}}
\label{fig:neffsptact}
\end{figure*}

In this section we explore the bounds on the $\neff$ parameter,
neglecting light neutrino masses and therefore assuming that there
exist in nature $\neff$ massless neutrino species. 
The left (right) panel of Fig.~\ref{fig:neffsptact} shows the
constraints in the ($\Omega_{\textrm{dm}}h^2$, $\neff$) plane arising
from the combination of WMAP plus SPT (ACT) as well as the other data
combinations shown in the previous sections. Notice that the mean value of $\neff$
is, in general, much higher in the case of the SPT data analyses. When considering
CMB data only, $\neff=3.93\pm 0.68$ for the case of WMAP plus SPT data,
while $\neff=2.74\pm 0.47$ if analysing WMAP and ACT data. The tension
among these two $\neff$ mean values gets diluted if BAO data and a
prior on $H_0$ from the HST experiment are added in the analyses. In
that case, $\neff=3.83\pm 0.41$ ($\neff=3.43\pm 0.36$) for WMAP plus
SPT (ACT), being these two measurements perfectly consistent and
indicating both $\neff >3$ at 1-2 standard deviations. The addition of
SNLS3 data will not help much in improving the agreement between these
two data sets, see Tabs.~\ref{tab:c5} and \ref{tab:c6}, where we summarise the mean
values and errors found for $\neff$ for the different data
combinations considered here. Therefore, as in the three massive neutrino case,  the consistency between ACT and SPT CMB results is greatly improved
if BAO and HST data are considered as well.

\begin{table*}
\begin{center}
\begin{tabular}{lcccccc}
\hline \hline
         &      W9+SPT & W9+SPT & W9+SPT & W9+SPT& W9+SPT & W9+SPT  \\
      &                      & + HST    &+BAO     &+SNLS3 & +BAO+HST&+BAO+SNLS3\\
\hline
\hspace{1mm}\\
${\neff}$ &  $3.93\pm0.68$ & $3.59\pm0.39$ & $3.50\pm0.59$ & $3.96\pm0.69$ & $3.83\pm0.41$ & $3.55\pm0.63$\\
\hspace{1mm}\\
\hline
\hline
\end{tabular}
\caption{Mean values and errors on ${\neff}$ in a standard cosmology with $\neff$ massless neutrinos for the
  different combinations of data sets in the case of considering high
  multipole data from SPT.}
\label{tab:c5}
\end{center}
\end{table*}

\begin{table*}
\begin{center}
\begin{tabular}{lcccccc}
\hline \hline
         &      W9+ACT & W9+ACT & W9+ACT & W9+ACT& W9+ACT & W9+ACT  \\
      &                      & + HST    &+BAO     &+SNLS3 & +BAO+HST&+BAO+SNLS3\\
\hline
\hspace{1mm}\\
${\neff}$ &  $2.74\pm0.47$ & $3.12\pm0.38$ & $2.77\pm0.49$ & $2.79\pm0.47$ & $3.43\pm0.36$ & $2.83\pm0.47$\\
\hspace{1mm}\\
\hline
\hline
\end{tabular}
\caption{Mean values and errors on ${\neff}$ in a standard cosmology with $\neff$ massless neutrinos for the
  different combinations of data sets in the case of considering high
  multipole data from ACT.}
\label{tab:c6}
\end{center}
\end{table*}

\subsection{Massive neutrinos and dark radiation}

\begin{figure*}
\begin{tabular}{c c}
\includegraphics[width=9cm]{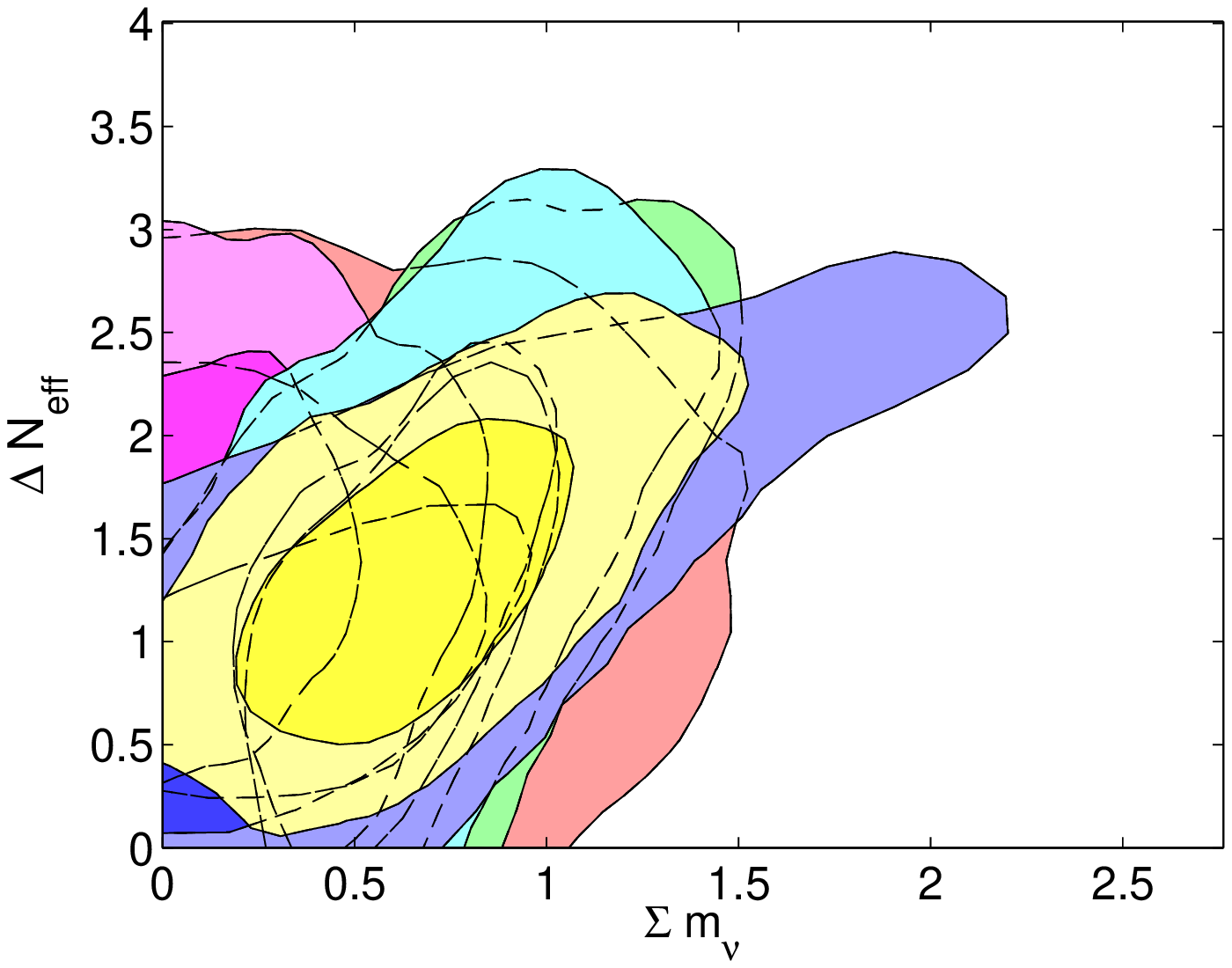}&\includegraphics[width=9cm]{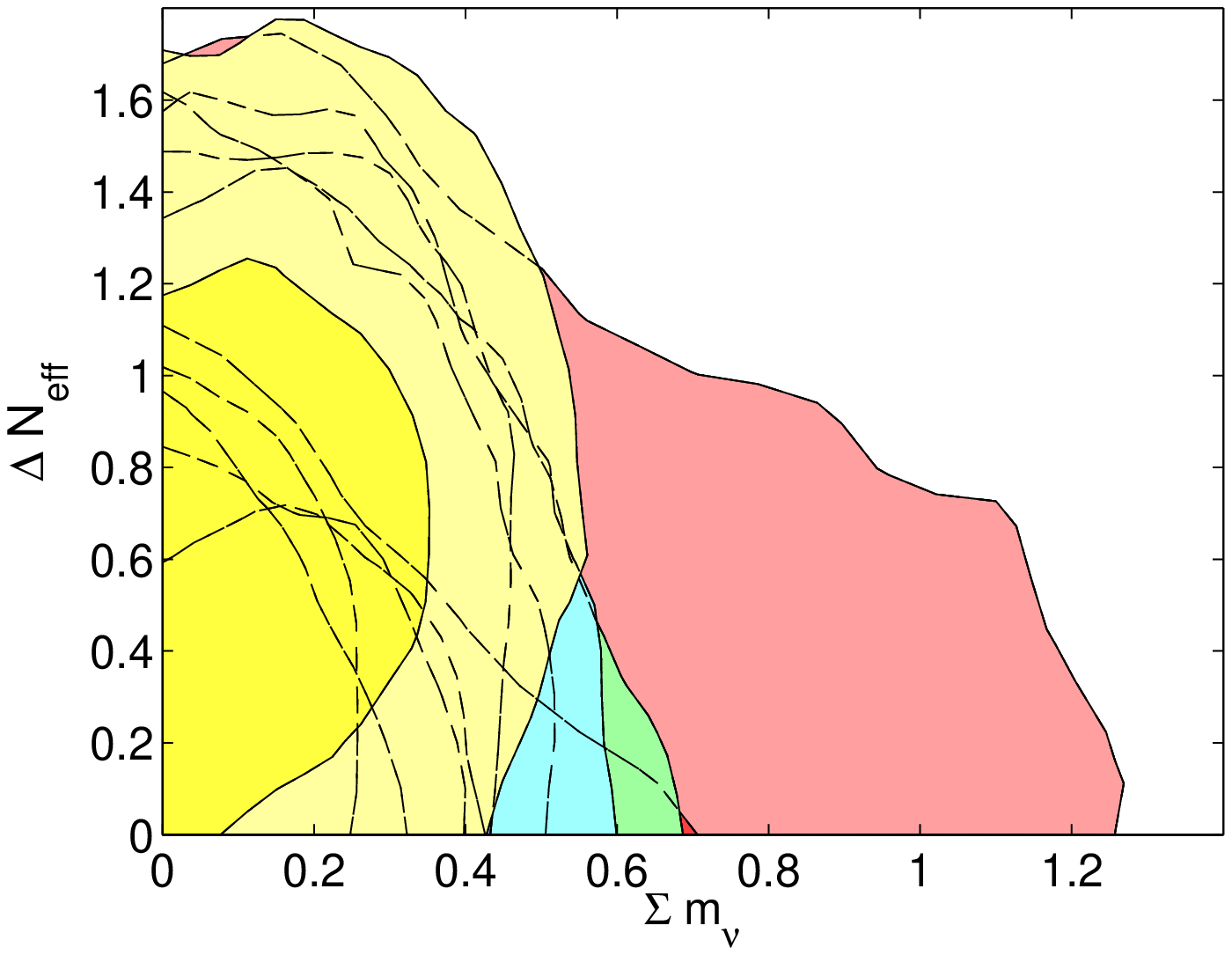}\\
\end{tabular}
 \caption{\small{Left panel ($\Delta \neff$ dark radiation species plus three
     massive neutrinos): the red contours show the $68\%$ and $95\%$~CL allowed
  regions  from the combination of
  WMAP and SPT measurements in the ($\sum m_\nu$ (eV), $\Delta \neff$)
  plane, while the magenta (blue) ones show the impact of the addition
  of SNLS3 (HST) data sets.  The green contours depict  the results
  from the combination of CMB and BAO data, while the cyan and  yellow ones show the impact of the SNLS3 (HST) data combined
  with CMB and BAO measurements. Right panel: as in the left panel but
  considering ACT measurements instead of SPT data.}}
\label{fig:case3}
\end{figure*}

\begin{figure*}
\begin{tabular}{c c}
\includegraphics[width=9cm]{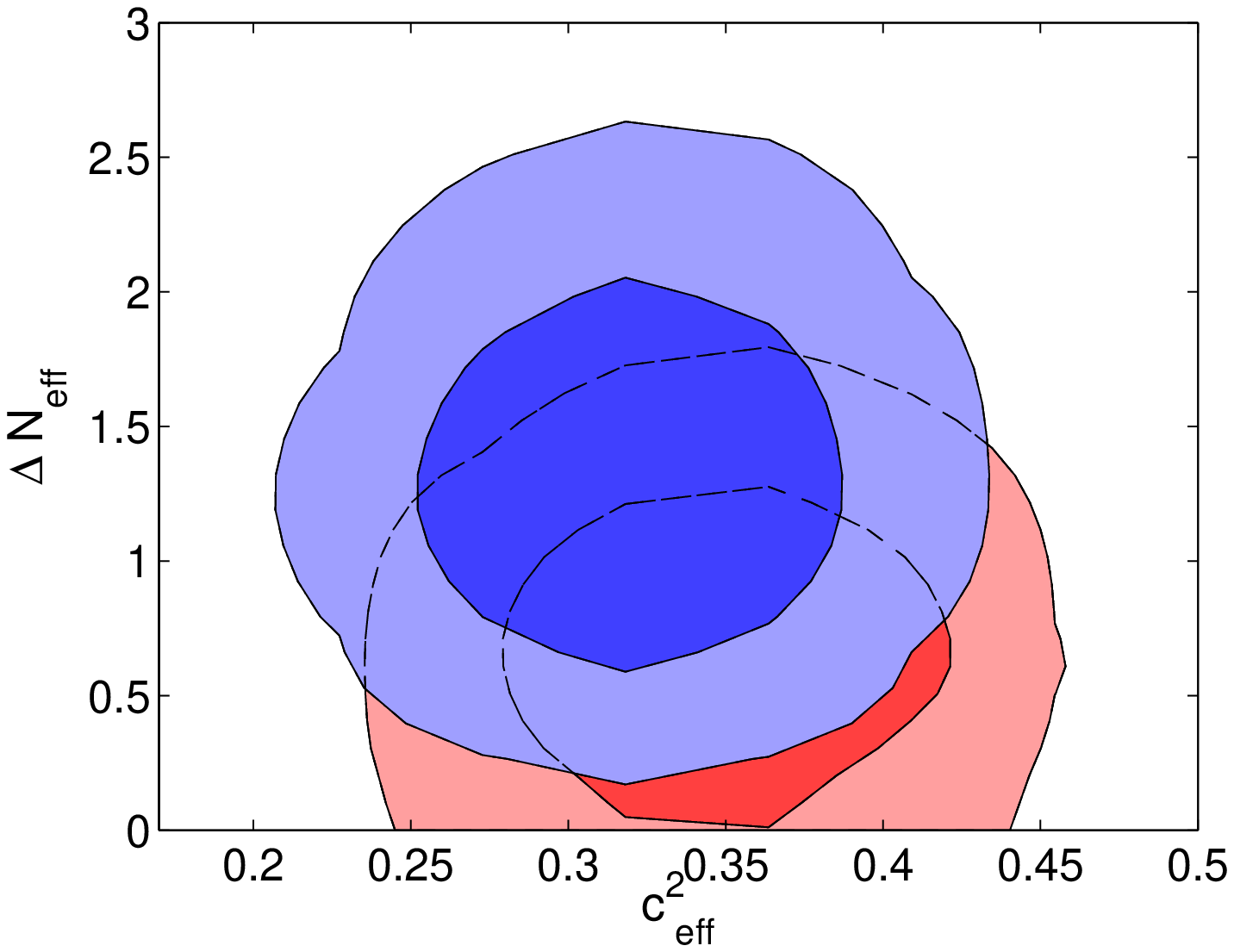}&\includegraphics[width=9cm]{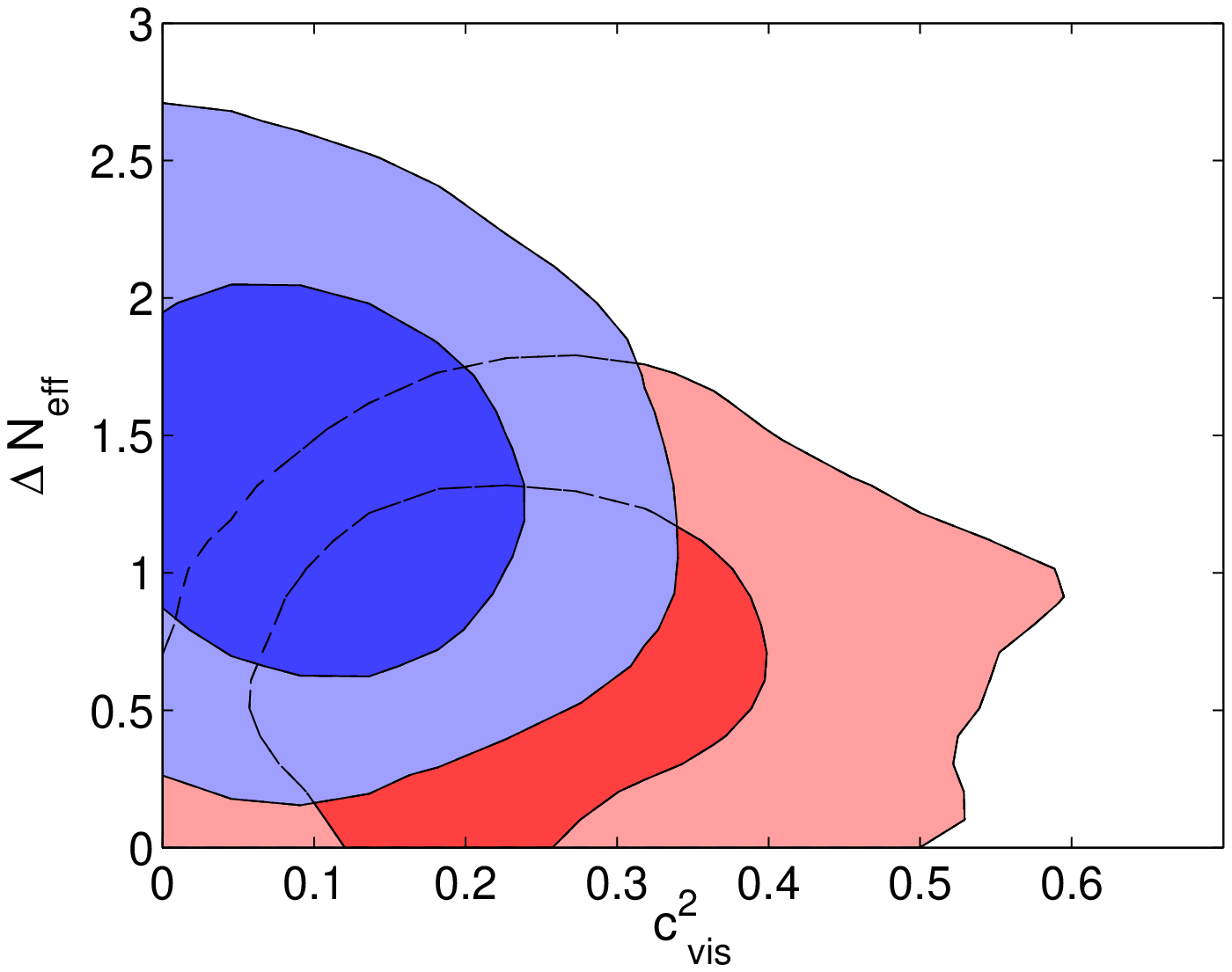}\\
\end{tabular}
 \caption{\small{Left panel ($\Delta \neff$ massless neutrinos and three
     massive with $\sum m_\nu=0.3$~eV): the red (blue) contours show the $68\%$ and $95\%$~CL allowed
  regions  from the combination of
  WMAP, SPT (ACT), BAO and HST measurements in the ($\ceff$, $\Delta
  \neff$) plane. Right panel: as in the left panel but in the ($\cvis$, $\Delta
  \neff$) plane.}}
\label{fig:case1}
\end{figure*}

In this section we consider extended dark radiation cosmologies,
parameterised via the dark radiation abundance $\neff$ and its
clustering properties, represented by  $\ceff$ and $\cvis$, see also
Refs.~\cite{Diamanti:2012tg,Archidiacono:2012gv,Archidiacono:2011gq}
for bounds on these parameters within different cosmological models.
Here three possible scenarios are examined. In the first scenario there are three massive
neutrinos with $\sum m_\nu=0.3$~eV, which roughly corresponds to the
mean value obtained in Ref.~\cite{Hou:2012xq},  and $\Delta \neff$ massless
neutrino species with $\cvis=\ceff=1/3$. In the second scenario  the clustering
parameters of the dark radiation component are allowed to vary, as
well as in the third scenario, in which also the sum pf the masses of the three massive neutrinos $\sum m_\nu$ varies. Our
findings are summarised in Figs.~\ref{fig:case1} and
\ref{fig:case3}, where we illustrate the constraints from SPT
and ACT probes. 

In the first scenario, in which both $\ceff$ and $\cvis$
are fixed to their standard values and assuming three massive neutrinos with $\sum
m_\nu=0.3$~eV, we find that $\Delta \neff=0.89\pm 0.56$ ($\Delta \neff=0.42\pm 0.34$) 
when considering WMAP plus SPT (ACT) measurements. 
When HST data are added in the analyses, the mean values
of $\Delta \neff$ for these two probes are similar: $\Delta \neff=0.95\pm 0.42$
($\Delta \neff=0.71\pm 0.40$) for WMAP, SPT and HST (WMAP, ACT and HST) measurements. 
 The addition of BAO data does not improve the agreement between SPT and ACT. 

In the second scenario, only $\sum m_\nu=0.3$~eV remains as a fixed
parameter. In this case, the discrepancy between SPT and ACT data sets
is larger, being the mean values for $\Delta \neff=1.31\pm 0.60$ and
$\Delta \neff=0.38\pm 0.32$ respectively. The addition of HST brings these two mean
values closer, being $\Delta \neff=0.92\pm 0.39$ ($\Delta \neff=0.62\pm 0.41)$ for
the combinations of WMAP, SPT and HST (WMAP, ACT and HST) data sets.
Concerning the values of the dark radiation clustering parameters $\ceff$ and
$\cvis$, we find that SPT data exclude the standard value of
$\cvis=1/3$ at the 2$\sigma$ CL. The
mean value is $\cvis=0.15\pm 0.07$ ($\cvis<0.28$ at $95\%$~CL) when
combining SPT, WMAP, BAO and HST data sets. 
The results for the effective speed of sound seem to be consistent
with standard expectations, finding, for the same combination of data
sets, that $\ceff=0.32\pm 0.012$. Similar results are obtained when SPT data is combined with the other data sets
considered here.  However, the results of the analyses of ACT data
provide values for the clustering parameters which perfectly agree with standard
expectations, being $\ceff=0.35\pm 0.02$ and $\cvis=0.25\pm 0.13$
($\cvis<0.61$ at $95\%$~CL) for the analysis of ACT, WMAP, BAO and HST
data sets.

In the third scenario, all four parameters $\Delta \neff$, $\cvis$,
$\ceff$ and $\sum m_\nu$ are allowed to freely vary, and we depict the
constraints arising from our analyses in Figs.~\ref{fig:case3}.
 The evidence for neutrino masses
when analysing SPT data gets diluted for all the data
combinations except when BAO data is also added in the analyses. 
We find $\Delta \neff=1.34\pm 0.67$, ($\Delta \neff=1.15\pm 0.64$)  and $\sum
m_\nu<1.3$~eV ($\sum m_\nu< 1.75$~eV) at $95\%$ CL when considering
CMB  (CMB plus HST) measurements.  For the combination of WMAP, SPT
and BAO (WMAP, SPT, BAO and HST) data sets, the cosmological evidence for neutrino masses still
remains, finding that $\Delta \neff=1.30\pm 0.77$ ($\Delta \neff=1.35\pm 0.50$) and $\sum
m_\nu=0.68\pm 0.31$~eV ($\sum m_\nu=0.67\pm 0.29$~eV). When analysing ACT data (see right panel of
Fig.~\ref{fig:case3}) the bounds on both $\Delta \neff$ and $\sum m_\nu$ are
tighter than those found for SPT data. For the combination of WMAP,
ACT, HST and BAO data sets, $\Delta \neff=0.74\pm 0.40$ and $\sum
m_\nu<0.46$~eV at $95\%$~CL. 
Regarding the values of $\ceff$ and $\cvis$, we find very similar results to those shown previously.
In this third scenario in which the sum of the three massive neutrinos
is also a free parameter, we find that SPT data again
exclude the standard value of $\cvis=1/3$ at the 2$\sigma$ CL,
while the value of $\ceff$ agrees with its standard prediction. The
analysis of SPT, WMAP, BAO and HST gives $\cvis=0.13\pm 0.07$
($\cvis<0.26$ at $95\%$~CL) and $\ceff=0.32\pm 0.01$. 
In the case of ACT data, the values for both clustering parameters perfectly agree with standard
expectations, being $\ceff=0.35\pm 0.02$ and $\cvis=0.25\pm 0.11$
($\cvis<0.47$ at $95\%$~CL) for the analysis of ACT, WMAP, BAO and HST
data sets.

Figure~\ref{fig:case1}, left (right) panel, shows the constraints on the dark
radiation abundance versus the effective speed of sound (viscosity
parameter) for the combination of SPT or ACT with WMAP, BAO and HST measurements.
Note that SPT and ACT data seem to be again in disagreement, this time
concerning the dark radiation clustering parameter $\cvis$.  

\section{Conclusions}
\label{sec:concl}

New Cosmic Microwave Background measurements have become recently available, motivating us to explore the
improvements in the measurements of the properties of the cosmic neutrino
and dark radiation backgrounds. Interestingly, the new measurements of
the CMB damping tail from the South Pole Telescope, SPT, and from the
Atacama Cosmology Telescope, ACT, seem to give different results concerning neutrino masses and abundances. While the SPT collaboration
finds $\sim 3\sigma$ evidence for neutrino masses and $\neff>3$ at
$\sim 2 \sigma$, the ACT collaboration does not find evidence for neutrino
masses and their value for $\neff$ is much lower, agreeing perfectly
with the standard model prediction of $\neff=3$. The success of future Majorana neutrino searches relies on
the absolute scale of neutrino masses; therefore a detailed
analysis of both data sets separately combined with other cosmological
measurements is mandatory. We have considered the most recent Baryon Acoustic
Oscillation data, measurements of the Hubble constant from the Hubble
Space Telescope, as well as Supernovae Ia luminosity distance
measurements. In the standard $\Lambda$CDM scenario with either three
massive neutrino species or $\neff$ massless species, the results from the two high CMB multiple probes
are consistent if Baryon Acoustic Oscillation data is removed from the
analyses and a prior on $H_0$ from HST is also considered. In the case
of $\neff$ massive neutrino species,  SPT and ACT data analyses give very different
results for $\sum m_\nu$: while the evidence for $\sum m_\nu \sim
0.5$~eV found for SPT data persists independently of the data sets combined,
the ACT data provide a $95\%$~CL upper bound of $\sim 0.4$~eV
on $\sum m_\nu$. We then explore extended cosmologies models, finding that, in
general, the SPT data evidence for neutrino masses found in the
minimal $\Lambda$CDM scenario gets diluted except for the case of a dark
radiation background of unknown clustering properties with BAO data
included. In the former case, SPT data exclude the standard value for
the viscosity parameter of the dark radiation fluid $\cvis=1/3$ at the
$2\sigma$ CL, regardless of the data sets considered in the analysis.
Upcoming, high precision CMB
data from the Planck satellite will help  in disentangling the high
tail CMB neutrino--dark radiation puzzle.

\section{Acknowledgments}
O.M. is supported by the Consolider Ingenio project CSD2007-00060, by PROMETEO/2009/116, by the Spanish Ministry Science project FPA2011-29678 and by the ITN Invisibles PITN-GA-2011-289442.
MA acknowledges the European ITN project Invisibles (FP7-PEOPLE-2011-ITN, PITN-GA-2011-289442-INVISIBLES).

%%%%%%%%%%%%%%%%%%%%%%%%%%%%%%%%%%%%%%%%%%%

%%%% new references

%%%%%%%%%%%%%%%%%%%%%%%%%%%%%%%%%%%%%%%%
%%%%%%%%%%%%%%%%%%%%%%%%%%%%%%%%%%%%%%%%

\end{document}